\documentclass[a4paper]{jpconf}
\usepackage{graphicx}

\usepackage{booktabs} 
\usepackage{array} 
\usepackage{verbatim} 
\usepackage{latexsym}
\usepackage{amssymb}
\usepackage{amsmath}

\begin{document}
\textsc{Prepared for submission to JPCS}

\title{On first attempts to reconcile quantum principles with gravity}
\author{A Rocci}
\address{Dipartimento di Fisica e Astronomia G. Galilei, Universit\`a degli Studi di Padova, I-35131 Padova (Italy)}
\ead{rocci@pd.infn.it}

\begin{abstract}
In his 1916's first paper on gravitational waves Einstein began to speculate on \textit{interactions} between the principles of the old quantum theory and  his theory of gravitation. With this contribution Einstein has stimulated a lot of similar speculations, during the dawn and the development of Quantum Mechanics. These speculations have culminated with the first attempt to quantize the gravitational field, that was provided by Rosenfeld in 1930. In this paper we briefly explain why this period (1916-1930) should be inserted into the history of Quantum Gravity and then we focus on Klein's approach to the problem of reconciling Wave Mechanics with gravity, during the two-years period 1926-1927. His attempt should be looked as the prehistory of Quantum Field Theory in a curved background.
\end{abstract}

\section{Introduction}

The term Quantum Gravity (QG) is often associated to the idea that the gravitational field must be quantized. Up to the present, we do not know how to construct this theory in a consistent way. From the birth of General Relativity (GR) in 1915 until today, many approaches tried to face, in a broad sense, the problem of harmonizing the quantum principles and GR. 

Einstein \cite{Einstein1} was the first to argue that quantum effects must modify his general theory in 1916, but what he had in mind was Bohr's principle of stationary orbits, that seems to be in conflict with the energy loss caused by gravitational wave emission. Presumably Einstein did not know, at that time, that this kind of atomic collapse is characterized by a time of the order of $10^{37}$s, that is an enormous lack of time compared with the life of our Universe\footnote{Recent results \cite{Planck} give approximately $(4.354 \pm 0.012)\times 10^{17}s$.} \cite{Gorelick1}.  

The first attempt to quantize the gravitational field appeared in 1930. In this year Rosenfeld published two papers. In his first paper\footnote{See \cite{Salisbury} for a commented English translation.} \cite{Rosenfeld1} Rosenfeld considers a Lagrangian that includes the electromagnetic, the matter (Dirac) and the gravitational fields. Concerning the latter, he writes the Einstein-Hilbert action using the tetrad formalism\footnote{Rosenfeld refers to this formalism as ``the one-body theory proposed by Weyl \cite{Weyl1}''.}. The quantization procedure takes place when he substitutes classical variables and their conjugate momenta with hermitian operators\footnote{In those years physicists were used to talk about \textit{c-numbers} (classical) and \textit{q-numbers} (quantum).}. With this paper the \textit{canonical quantization approach} enters into the history of QG. In a second paper \cite{Rosenfeld2} Rosenfeld uses the linearization procedure for the metric $g_{\mu\nu}$, introduced by Einstein\footnote{We use the signature $ (-,+,+,+) $ and $ \mu ,\nu $ run from 0 to 3.}\cite{Einstein1}: $g_{\mu\nu}=\eta_{\mu\nu}+\sqrt{\chi}h_{\mu\nu}$, where $\chi$ is proportional to the Newton constant and $h_{\mu\nu}$ is the perturbation of the Minkowski metric $\eta_{\mu\nu}$. In this article Rosenfeld writes the linearized Lagrangian and chooses, as field variables, the perturbation of the metric, that he rewrites in terms of annihilation and creation operators. This splitting procedure will be adopted also later on by Gupta \cite{Gupta} and Feynman \cite{Feynman}, the fathers of the \textit{covariant quantization approach}.

Now we want to raise the following question: should the attempts \textit{preceding} Rosenfeld's work be considered part of the history of QG?  To answer the question we point out the following facts. First of all, the idea that \textit{every field can be quantized} appeared for the first time in 1929, thanks to Heisenberg and Pauli \cite{Heisenberg-Pauli}. Despite of this fact, Einstein started before to speculate on possible interactions between the principles of the old quantum theory\footnote{This term refers, as usual, to all attempts to explain new phenomena that arose between 1900 and 1925, using quantized classical quantities.} and his theory of gravitation. Likewise, Klein\footnote{We refer to Oskar Klein, not to be confused with Felix Klein.} and other researchers, during the early years of Quantum Mechanics (QM), tried to find out a unifying framework that could harmonize Wave Mechanics\footnote{In the old quantum theory there were two main approaches \cite{Pais2}: the \textit{particle} approach and the \textit{wave} approach. They led to the birth of Matrix Mechanics in 1925, developed by Heisenberg, Born and Jordan, and to the birth of Wave Mechanics at the beginning of 1926, developed by Schr\"odinger, respectively.} with gravity. It was a failed\footnote{Klein himself admitted that his program failed because ``as Dirac may well say, my main trouble came from trying to solve too many problems at a time!''\cite{Klein-Life}.} attempt to modify GR for reconciling it with quantum principles and some authors have already underlined this attempt \cite{liv-rev} \cite{Raife1} \cite{Raife2}. For this reason we decided to study more in details Klein's program from the perspective of the history of QG.
To answer the question we posed, another important fact to consider is the role played by the semi-classical methods born after Rosenfeld papers, like e.g. Quantum Field Theory (QFT) in curved space-times. Even though they are not directly connected with QG problems, they revealed new phenomena and raised new problems that will find a solution in a consistent theory of QG. We have in mind the black hole radiation \cite{Hawking} and the \textit{information loss paradox} \cite{LossParadox}, discovered by Hawking in the seventies. Concerning the latter, it will be avoided by a unitary description of the black hole radiation process in the framework of a consistent theory of QG. From a modern point of view, QFT in curved backgrounds is a \textit{semi-classical} approach, because \textit{quantum} matter fields live in a \textit{classical} curved background: we can consider this approach as an attempt to reconcile quantum principles with gravity at an energy scale where the quantum effects of the gravitational field are negligible and for this reason we believe that the history of these methods should belong to the history of QG. And last but not least, let consider String Theory, one of the most promising approach to Q. It combines together various features coming from different areas, like e.g. Klein's idea of extra dimensions or QFT in two dimensions, without quantizing the gravitational field directly. In fact String Theory offers a picture where QG \textit{emerges} from the quantum theory of strings \cite{Polchinski}.

For these reasons we think that \textit{every} attempt to reconcile quantum principles with gravity should belong to the history of QG, even though the attempt does not try to quantize the gravitational field. We imagine the history of QG as an island chain; on every island there is a farm, with physicists at work, and every island corresponds to a different approach to the problem of reconciling quantum postulates with the theory of gravitation.

Our paper aims to consider the five-dimensional farm in the years 1926 and 1927, with the main focus on Klein's work. The five-dimensional farm was an active research program that tried to unify gravity, Maxwell's theory and Wave Mechanics, using a five-dimensional space-time as we will explain in detail in the following section. As far as we know, we note for the first time that Klein's contribution should be inserted into the \textit{prehistory} of QFT in curved backgrounds.

\section{The Klein's contribution}
Maxwell's legacy has shown that unifying different areas of physics could lead to a better understanding of our world. Following this idea, some researchers started to search a \textit{unified theory}, that could describe both GR and Maxwell's theory in a unified framework. Among these, some physicists assumed that we live in a \textit{five-dimensional} world: in addition to the usual four space-time dimensions, there is a fifth space-like dimension\footnote{Because of this assumption, from now on, we will call this approach \textit{the five-dimensional farm}.}, $ x^5 $, that is compactified on a circle\footnote{This Ansatz is often known as the hypothesis of \textit{cilindricity}.} of radius $ l $. In this approach the five-dimensional metric, $ \gamma_{\bar{\mu}\bar{\nu}} \quad \bar{\mu},\bar{\nu} = 0,1,2,3,5$, once decomposed in its low-dimensional components, can describe gravity and Maxwell's theory. In fact $ \gamma_{\mu\nu} $ and $ \gamma_{5\mu} $ behave like a four-dimensional symmetric tensor and a four-dimensional vector respectively and then they could play the role of the metric and of the electromagnetic potential. Kaluza and Klein are known as the fathers of this approach\footnote{The $ \gamma_{55} $ component behaves like a scalar field and it is today known as the \textit{dilaton}. Its role was not fully understood at that time: it will be set to be a constant by these authors.}. 

Kaluza made the assumption of cilindricity in 1921 \cite{Kaluza}. In the  last sentence of his paper he declares to be afraid that quantum theory would be a menace for his theory\footnote{``After all, what threatens all the Ansatz, which demand universal validity, is the sphinx of modern physics - quantum theory''}. Klein found out the same approach later and independently in 1926. On the contrary he was convinced that quantum principles should play a fundamental role in his version of a five-dimensional unified theory. 

Klein's program can be summarized in the following three steps: writing a five-dimensional wave equation\footnote{As we will see, the wave equation he used is the \textit{Klein-Gordon} (K-G) \textit{equation}.} for a massive particle (electron/proton); embedding it into a curved space-time; getting a conservation principle. 
In his first paper \cite{Klein1}, he introduces two five-dimensional metrics, related to each other and depending on the four-dimensional $ x^{\mu} $ only.
The first line element is $d\sigma^2 = \gamma_{\bar{\mu}\bar{\nu}}(x^{\mu})dx^{\bar{\mu}}dx^{\bar{\nu}} $ that he decomposes in the following way:
\begin{equation}
\begin{split}
d\sigma^2 & = \gamma_{55} (dx^5)^2 + 2\gamma_{5\mu}dx^5dx^{\mu} + \gamma_{\mu\nu}dx^{\mu}dx^{\nu} 
 = \gamma_{55}\left( dx^5 + \frac{\gamma_{5\mu}}{\gamma_{55}}dx^{\mu} \right) ^2+ \left( \gamma_{\mu\nu} - \frac{\gamma_{5\mu}\gamma_{5\nu}}{\gamma_{55}} \right) dx^{\mu}dx^{\nu} \\
& = \alpha d\vartheta ^2 + g_{\mu\nu}dx^{\mu}dx^{\nu} = \alpha d\vartheta ^2 + ds^2. 
\end{split}
\end{equation}
Klein decided to set $\alpha = \frac{16\pi G}{e^2c^2} $, where $G,c$ and $e$ are the Newton constant, the speed of light and the electric charge respectively. Using these Ansatz, Klein proved that the variational principle applied to an Einstein-Hilbert action in five dimensions leads to a set of equation that can be identified with four-dimensional Einstein's equations and the Maxwell equations\footnote{In this sense, Klein's theory unifies gravitational and electromagnetic forces. $T^{\mu\nu}$ is the usual electromagnetic energy momentum tensor.}:
$$
\delta\int d^5x \sqrt{-\gamma} R^{(5)} = 0 \quad\rightarrow\quad {R^{(4)}}^{\mu\nu}-\frac{1}{2}g^{\mu\nu}R^{(4)} = \frac{8\pi G}{c^4} T^{\mu\nu}  \quad , \quad\partial_{\mu}\left( \sqrt{-g}F^{\mu\nu} \right) = 0
$$
This happens because Klein identifies the $g_{\mu\nu}$ of eq. (1) with the four-dimensional metric and because he defines the electromagnetic four-potential $ A_{\mu} $, whose field strength is $ F^{\mu\nu} $, in the following way: 
$$ A_{\mu}=\frac{c}{e}\left( \frac{\gamma_{5\mu}}{\gamma_{55}}\right) \qquad F_{\mu\nu}=\partial_{\mu}A_{\nu}-\partial_{\nu}A_{\mu}. $$

The second line element that he introduces is $ d\hat{\sigma}^2 = a_{\bar{\mu}\bar{\nu}}dx^{\bar{\mu}}dx^{\bar{\nu}} $ defined by:
$$
 d\hat{\sigma}^2 = \hat{k} d\vartheta ^2 + ds^2 = \gamma_{\bar{\mu}\bar{\nu}}dx^{\bar{\mu}}dx^{\bar{\nu}} + \left( \hat{k}-\alpha\right)  d\vartheta ^2 = d\sigma^2 + \left( \hat{k}-\alpha\right)  d\vartheta ^2  
$$
After setting $\hat{k} = \frac{1}{m^2c^2}$, where $m$ represents the electron mass, Klein writes his wave equation\footnote{For scalar functions one covariant derivative is equivalent to an ordinary derivative.}:
\begin{equation}
a^{\bar{\mu}\bar{\nu}}\left( \frac{\partial}{\partial x^{\bar{\mu}}} +\Gamma_{\bar{\mu}\bar{\nu}}^{\bar{\sigma}}\right)\partial_{\bar{\sigma}}\Psi= a^{\bar{\mu}\bar{\nu}}D_{\bar{\mu}}\partial_{\bar{\nu}}\Psi=0. 
\end{equation}
The \textit{covariant derivative} $ D_{\bar{\mu}} $ tells us that the dynamic is on a curved manifold. Eq. (2) shows that he that he introduced the second metric in order to write a \textit{massless} wave equation\footnote{The mass is hidden into the metric $a_{\bar{\mu}\bar{\nu}}$.}, in analogy with light. From a modern point of view, this is the wave equation for a massless scalar field. Using the same metric and following the program sketched above, Klein introduces a Lagrangian for a \textit{massless} particle in five-dimension:
$$ L = \frac{1}{2}a_{\bar{\mu}\bar{\nu}}\frac{dx^{\bar{\mu}}}{d\lambda}\frac{dx^{\bar{\nu}}}{d\lambda} = \frac{1}{2}\left[ \hat{k} \left( \frac{d\vartheta}{d\lambda}\right)^2 + 
\left( \frac{ds}{d\lambda}\right)^2\right]  $$
where $ \lambda $ is the affine parameter that parametrizes the geodesics.
The conjugated momenta $ p^{\bar{\mu}} $, defined as usual by
$ \displaystyle{p^{\bar{\mu}}=
\frac{\partial L}{\partial (dx^{\bar{\mu}} / d\lambda)}}$, produce a conservation principle, $ \displaystyle{\frac{dp^5}{d\lambda}} = 0 $ because the metric does not depend on $ x^5 $, and the four-dimensional geodesic equation for a \textit{charged electron} moving in a gravitational and electromagnetic field.

The unusual connection between the \textit{massless} wave equation and the \textit{massive} electron was noted by de Broglie at the beginning of 1927 \cite{deBroglie} and in his last paper \cite{Klein2} Klein rewrites eq. (2) in an equivalent form, using the inverse of the \textit{first} metric we introduced, $\gamma^{\bar{\mu}\bar{\nu}} $:
\begin{equation}
\gamma^{\bar{\mu}\bar{\nu}}D_{\bar{\mu}}\partial_{\bar{\nu}}\Psi=\mathcal{I}^2\Psi \qquad \mathcal{I} ^2=\frac{1}{\hbar ^2}\left( m^2c^2-\frac{e^2c^2}{16\pi G}\right).
\end{equation} 
From our point of view, this is the K-G equation for a \textit{massive} scalar field living in a curved background. 

In particular, in this equation it appears the Planck constant\footnote{De Broglie was proud of the fact that this equation contains \textit{all} known constants of Nature.} $ \hbar $. To understand this fact, we start noting that the momentum $ p^5 $ does not have the usual dimensions, because $ \alpha $, and then $ x^5 $, is not dimensionless\footnote{See equation (1)}. This means that the fifth coordinate having the right length dimensions is $ \tilde{x}^5 = \sqrt{\alpha}x^5 $. Relaxing completely the analogy with light, Klein introduces another Lagrangian \cite{KleinNature} using the \textit{first} metric: 
$$\tilde{L} = \frac{1}{2}m\gamma_{\bar{\mu}\bar{\nu}}\frac{dx^{\bar{\mu}}}{d\lambda}\frac{dx^{\bar{\nu}}}{d\lambda}.$$
This is the Lagrangian for a \textit{massive} particle in five dimensions and the fifth component of the momentum associated to $ \tilde{x}^5 $  is conserved again: $\displaystyle{\frac{d\tilde{p}^5}{d\lambda} = 0}$. This means that $ \tilde{p}^5 $ is a constant to be determined. As before, if we ask to the four-momentum $ p^{\mu} $ to reproduce the geodesic equation, the following identity must be satisfied: 
$ \sqrt{\alpha}\tilde{p}^5 = 1 $. Using the Ansatz of cilindricity, Klein writes $ \tilde{p}^5 = \frac{nh}{l}$, as usual in QM, and he sets $ n=1 $. The wave function $ \Psi $ must be periodic in the fifth coordinate and equating both expression for $ \tilde{p}^5 $ we get the period $ l $: $ l = \sqrt{\alpha} h $. Inserting this conditions into eq. (2) makes the Planck constant to appear. From Klein's point of view, these equalities give a connection between the quantization rules and the quantization of the electric charge. But the only consequence is an estimation for the radius of the fifth dimension: the Planck length. Klein did not consider that setting $ n=1 $ implies to have a particle with mass equal to the Planck mass, as it happens today in String Theory \cite{Polchinski}.

In his last paper of the two-year period \cite{Klein2}, Klein follows his program again, but looking for a more general approach. In fact the starting point is the following invariant, that is proportional to the action for a massive scalar field\footnote{The action did not have the meaning that we give it today in field theory and an overall minus sign is missing in front of the action.}:
\begin{equation}
\mathcal{S} =  \int \sqrt{-\gamma}\,\mathcal{L}\, d^5x = \int \sqrt{-\gamma}\,\frac{\hbar ^2}{m}\left[\frac{1}{2}\partial_{\bar{\mu}}\Psi\partial^{\bar{\mu}}\Psi+\frac{1}{2}\mathcal{I}^2\Psi^2\right]\, d^5x .
\end{equation}
Now, varying the action with respect to the dynamical variables $ \Psi $ and $ \gamma_{\bar{\mu}\bar{\nu}} $, he obtains a new five-dimensional \textit{conservation principle}. In fact we can write: 
$$
\delta\mathcal{S} = \delta_{\gamma}\mathcal{S} + \delta_{\Psi}\mathcal{S}
= \int \sqrt{-\gamma}\,\left[ \Theta_{\bar{\mu}\bar{\nu}}\delta \gamma^{\bar{\mu}\bar{\nu}}+ \frac{\hbar ^2}{m}\left( D_{\bar{\mu}}\partial^{\bar{\mu}}\Psi -\mathcal{I}^2\Psi\right)\delta\Psi\right] d^5x 
$$
where $\Theta_{\bar{\mu}\bar{\nu}}$ is one half of the energy-momentum tensor of the scalar field as we define it today \cite{DeFelice}. Introducing an infinitesimal variation of the  coordinates $x^{\bar{\mu}}\rightarrow x^{\bar{\mu}}+\xi^{\bar{\mu}}$ the fields vary as usual, $\delta\Psi = \partial_{\bar{\mu}}\Psi\xi^{\bar{\mu}}\quad \delta\gamma_{\bar{\mu}\bar{\nu}} = D_{(\bar{\mu}}\xi_{\bar{\nu})} $, and integrating by parts we get:
$$
\delta\mathcal{S} =\int \sqrt{-\gamma}\,\left[ D^{\bar{\mu}}\Theta_{\bar{\mu}\bar{\nu}} + \frac{\hbar ^2}{m}\left( D_{\bar{\mu}}\partial^{\bar{\mu}}\Psi - \mathcal{I}^2\Psi\right) \partial_{\bar{\nu}}\Psi\right] \xi^{\bar{\nu}} d^5x
$$

At the end, imposing the variational principle and the wave equation for $\Psi$ in the last equation, Klein obtains the covariant version of the energy-momentum conservation law\footnote{He also applies this procedure to the free electromagnetic field case, where it works in a similar way.}: $ D^{\bar{\mu}}\Theta_{\bar{\mu}\bar{\nu}} = 0 $. 
\newline
We would like to conclude this section with some  observations.

As far as we know the action for a scalar field, the related energy-momentum tensor and its conservation law, connected with the free equations, appear for the \textit{first} time in Klein's papers. Klein never used the word \textit{field} referring to the $ \Psi $, he always explicitly considers it as the wave function of the electron. In spite of this, his approach is similar to the point of view that will be adopted by developers of QFT in curved backgrounds.

In 1927 Klein and Jordan \cite{Jordan-Klein} published the paper where they connect the Bose-Einstein statistic to the K-G equation. Despite of this fact, at that time Klein considered $ \Psi $ like the wave function associated to the electron. Pais \cite{Pais} pointed out that, in the Jordan-Klein paper, the authors treated the wave function as a field, in fact they write $ \Psi $ in terms of annihilation and creation operator. Despite of this fact, in his subsequent paper on five-dimensional approach, the last we commented, Klein never use this decomposition.

We can appreciate how Klein changed his point of view, during the two-year period. In the Concluding Remarks of the first paper of the period \cite{Klein1}, he presents as fundamental his introduction of the fifth dimension. In particular he states that it is possible to understand this radical modification ``through quantum theory''. At the end of the introduction of his last paper \cite{Klein2}, Klein realizes that his approach is incomplete, but he is still convinced that his procedure is a natural starting point to construct a general theory of quantum fields\footnote{``Hierdurch und nach dem erw\"ahnten Gesichtpunkt scheint sich diese f\"unfdimensionale Form der Relativit\"atstheorie als der nat\"urlische Ausgangspunkt f\"ur eine allgemeine Quantenfeldtheorie darzubieten''.}.

\section{Conclusions}
In the years 1926-1927 the five-dimensional farm was an active research area. All physicists belonging to the farm tried to modify GR to reconcile it with Wave Mechanics. From our modern point of view they lay in an intermediate stage between the first and the second quantization approach. In Klein's papers we find for the first time the action, the equations of motion and the energy-momentum tensor for a massive scalar field living in a curved background. His starting-point will be shared by developers of QFT in curved backgrounds. Even though they are not directly connected with QG, these attempts should belong to the history of QG. In particular we suggest to insert Klein's contribution in the \textit{prehistory} of QFT in curved backgrounds.

\section*{Acknowledgments}
We are grateful to Prof. K. Lechner and Prof. G. Peruzzi for the patience they always demonstrate, for fruitful discussions and for their valuable suggestions. We are definitely indebted to Sara for her loving and fundamental support.

\end{document}